\PassOptionsToPackage{english}{babel}

\documentclass[%
11pt,
aip,
amsmath,amssymb,
preprint,
floatfix
]{revtex4-1}

\usepackage{graphicx}
\usepackage[mathlines]{lineno}
\usepackage[english]{babel}
\usepackage{siunitx}
\usepackage[utf8]{inputenc}
\usepackage{marginnote}

\newcommand{\D}[1]{\ensuremath{\operatorname{D}\!{#1}}}

\def\pct{\%}
\DeclareSIUnit\centimeter{cm}

\begin{document}

\selectlanguage{English}

\title{Observation and quantification of inertial effects on the
drift of floating objects at the ocean surface} 

\author{M.\ J.\ Olascoaga} \email{jolascoaga@miami.edu}
\affiliation{Department of Ocean Sciences,  Rosenstiel School of
Marine and Atmospheric Science, University of Miami, Miami, Florida
33149, USA}

\author{F.\ J.\ Beron-Vera} \affiliation{Department of Atmospheric
Sciences, Rosenstiel School of Marine and Atmospheric Science,
University of Miami, Miami, Florida 33149, USA}

\author{ P.\ Miron} \affiliation{Department of Atmospheric Sciences,
Rosenstiel School of Marine and Atmospheric Science, University of
Miami, Miami, Florida 33149, USA}

\author{J.\ Tri\~nanes} \affiliation{Instituto de Investigaci\'ones
Tecnol\'oxicas, Universidade de Santiago de Compostela, Santiago,
Spain.} \affiliation{Cooperative Institute for Marine and Atmospheric
Studies, Rosenstiel School of  Marine and Atmospheric Science,
University of Miami, Miami, Florida 33149} \affiliation{Physical
Oceanography Department, Atlantic Ocean and Meteorological Laboratory,
National Oceanic and Atmospheric Administration, Miami, Florida
33149, USA}

\author{N.\ F.\ Putman} \affiliation{LGL Ecological Research
Associates, Inc., Bryan, Texas 77801, USA} 

\author{R.\ Lumpkin} \affiliation{Physical Oceanography Department,
Atlantic Ocean and Meteorological Laboratory, National Oceanic and
Atmospheric Administration, Miami, Florida 33149, USA}

\author{G.\ J.\ Goni} \affiliation{Physical Oceanography Department,
Atlantic Ocean and Meteorological Laboratory, National Oceanic and
Atmospheric Administration, Miami, Florida 33149, USA}

\date{Started: July 14, 2019. This version: \today.}

\begin{abstract}
  We present results from an experiment designed to better understand
  the mechanism by which ocean currents and winds control flotsam
  drift.  The experiment consisted in deploying in the Florida
  Current and subsequently satellite tracking specially designed
  drifting buoys of varied sizes, buoyancies, and shapes.  We explain
  the differences in the trajectories described by the special
  drifters as a result of their inertia, primarily buoyancy, which
  constrains the ability of the drifters to adapt their velocities
  to instantaneous changes in the ocean current and wind that define
  the carrying flow field.  Our explanation of the observed behavior
  follows from the application of a recently proposed Maxey--Riley
  theory for the motion of finite-size particles floating at the
  surface ocean.  The nature of the carrying flow and the domain
  of validity of the theory are clarified, and a closure proposal
  is made to fully determine its parameters in terms of the carrying
  fluid system properties and inertial particle characteristics.
\end{abstract}


\maketitle

\section{Introduction}

The assessment of motions of floating matter in the ocean is of
importance for a number of key reasons.  These range from improving
search-and-rescue operations at sea \cite{Breivik-etal-13,
Bellomo-etal-15};  to better understanding the drift of flotsam of
different nature including macroalgae such as \emph{Sargassum}
\cite{Gower-King-08, Brooks-etal-19, Wang-etal-19}, plastic litter
\cite{Law-etal-10, Cozar-etal-14}, airplane wreckage
\cite{Trinanes-etal-16, Miron-etal-19b}, tsunami debris
\cite{Rypina-etal-13a, Matthews-etal-17}, sea-ice pieces
\cite{Szanyi-etal-16}, larvae \cite{Paris-etal-20, Putman-etal-16},
and oil \cite{Olascoaga-Haller-12, Gough-etal-19}; to better
interpreting ``Lagrangian'' observations in the ocean
\cite{Lumpkin-Pazos-07, Beron-etal-16}.  At present, largely
piecemeal, ad-hoc approaches are taken to simulate the effects of
ocean currents and winds on the drift of floating objects.  A
systematic approach ideally founded on first principles is needed.
In an effort to contribute to building one, several experiments that
involved the deployment and subsequent satellite tracking of specially
designed drifting buoys of varied sizes, buoyancies, and shapes
were carried out in the North Atlantic.

In this work we report results of the first experiment in the Florida
Current.  The drifters were deployed at once in coincidental position,
off the southeastern coast of the Florida Peninsula.  The differences
in their trajectories are here explained as resulting from
\emph{inertial} effects, i.e., those due to the buoyancy and finite
size of the drifters, which prevent them from instantaneously
adjusting their velocities to changes in the carrying ocean current
and wind fields.  This is done by making use of a recently proposed
framework for surface ocean inertial particle motion \cite{Beron-etal-19b},
which is derived from the Maxey--Riley set \cite{Maxey-Riley-83},
the de-jure framework for the study of inertial particle dynamics
in fluid mechanics \cite{Michaelides-97, Provenzale-99,
Cartwright-etal-10}.

The standard Maxey--Riley set \cite{Maxey-Riley-83} is a classical
mechanics second Newton's law that approximates the motion of
inertial particles immersed in a fluid in motion.   As such, it is
given in the form of an ordinary differential equation, rather than
a partial differential equation that would result from the exact
formulation of the motion, which involves solving the Navier--Stokes
equation with a moving boundary.  The latter is a formidable task
which would hardly provide as much insight as the analysis of an
ordinary differential equation can provide.

The type of insight that analysis of the Maxey--Riley set can lead
to includes foundation for realizing that the motion of neutrally
buoyant particles should not synchronize with that of fluid particles,
irrespective of how small \cite{Babiano-etal-00, Vilela-etal-06,
Sapsis-Haller-08}.  Additional insight includes that which followed
from earlier geophysical adaptions of the Maxey--Riley set, to wit,
the possible role of mesoscale eddies as attractors of inertial
particles \cite{Beron-etal-15, Haller-etal-16} and the tendency of
the latter to develop large patches in the centers of the subtropical
gyres \cite{Beron-etal-16}.

It is important to stress that the Maxey--Riley modeling framework
for inertial particle motion on the ocean surface \cite{Beron-etal-19b}
is quite different than the so-called leeway modeling approach of
search-and-rescue applications at sea \cite{Breivik-Allen-08}.  In
such an approach, widely used for its simplicity \cite{Duhec-etal-15,
Trinanes-etal-16, Allshouse-etal-17}, windage effects on objects
are modeled by means of a velocity resulting from the addition of
a small fraction of the wind field, established in an ad-hoc manner,
to the surface ocean velocity

The rest of the paper is organized as follows. Section~\ref{sec:fe}
describes the field experiment.  The Maxey--Riley set for inertial
ocean particle dynamics derived by \citet{Beron-etal-19b} is presented
in Section~\ref{sec:mrf} and clarified in Section~\ref{sec:mrclar}
with respect to the nature of the carrying flow, its domain of
validity, and parameter specification.  Section~\ref{sec:mrexp}
describes the application of the Maxey--Riley framework to explain
the behavior of each drifter type during the field experiment.
Finally, Section~\ref{sec:con} offers a summary and the conclusions
of the paper.

\section{The field experiment}\label{sec:fe}

The field experiment consisted in deploying simultaneously objects
of varied sizes, buoyancies, and shapes on 7 December 2017 at
(79.88$^{\circ}$W, 25.74$^{\circ}$N), situated off the southeastern
Florida Peninsula in the Florida Current, and subsequently tracking
them via satellite.  These buoys will be referred to as \emph{special
drifters} to distinguish them from other more standardized drifter
designs such as those from the Global Drifter Program (GDP).  The
special drifters were designed at the National Oceanic and Atmospheric
Administration's Atlantic Oceanographic and Meteorological Laboratory
for this experiment. 

\begin{figure}[t!]
  \centering%
  \includegraphics[width=.75\textwidth]{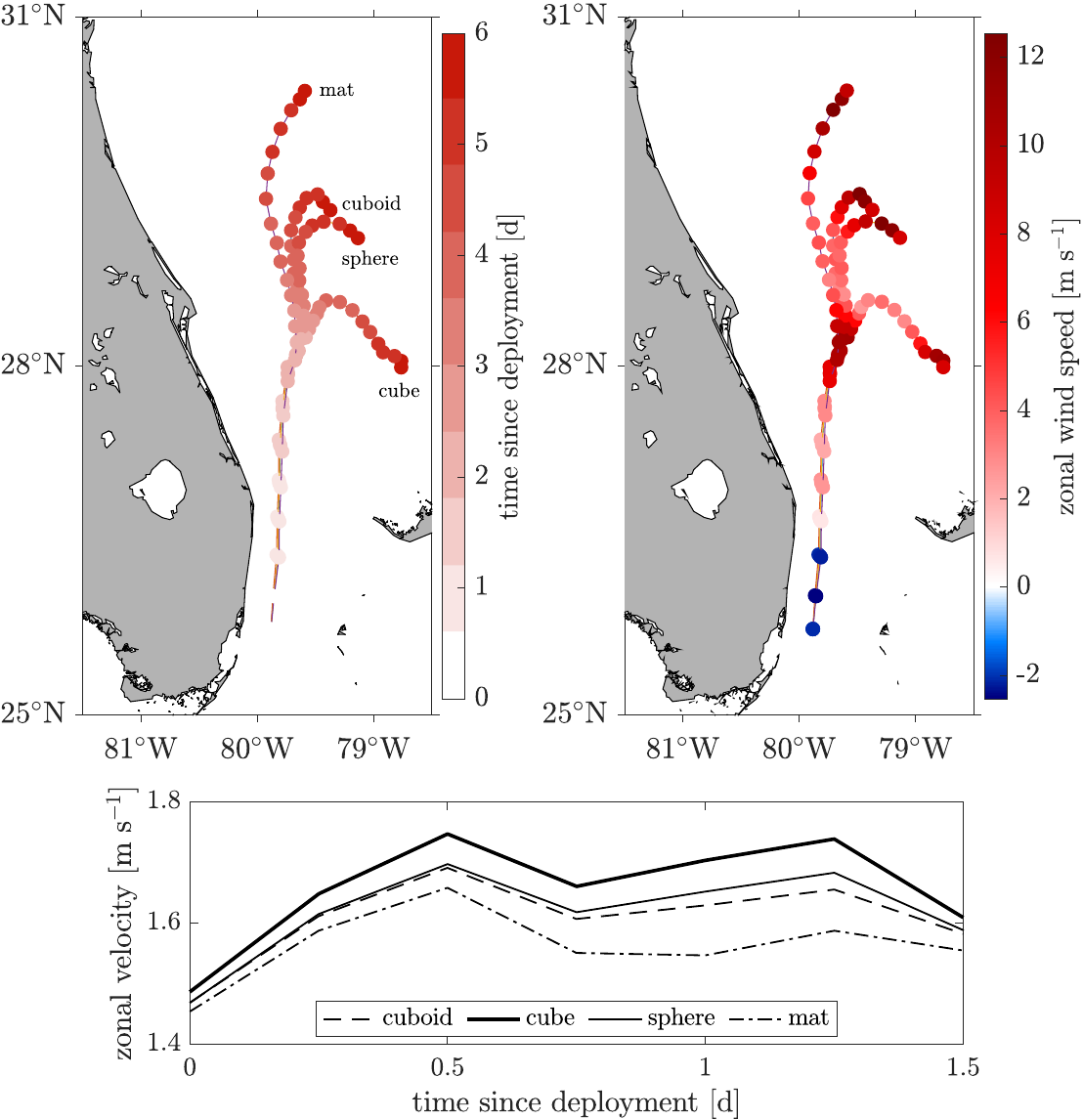}%
  \caption{(top) Satellite-tracked trajectories of the special
  drifters with colors indicating time since deployment (left) and
  zonal (i.e., west-to-east) wind intensity (right).  (bottom) Zonal
  velocity of the special drifters as a function of time from
  deployment to the instant when zonal wind speed reached its first
  peak.}
  \label{fig:time-wind}%
\end{figure}

\begin{figure}[t!]
  \linespread{1}\selectfont{}
  \centering%
  \includegraphics[width=.5\textwidth]{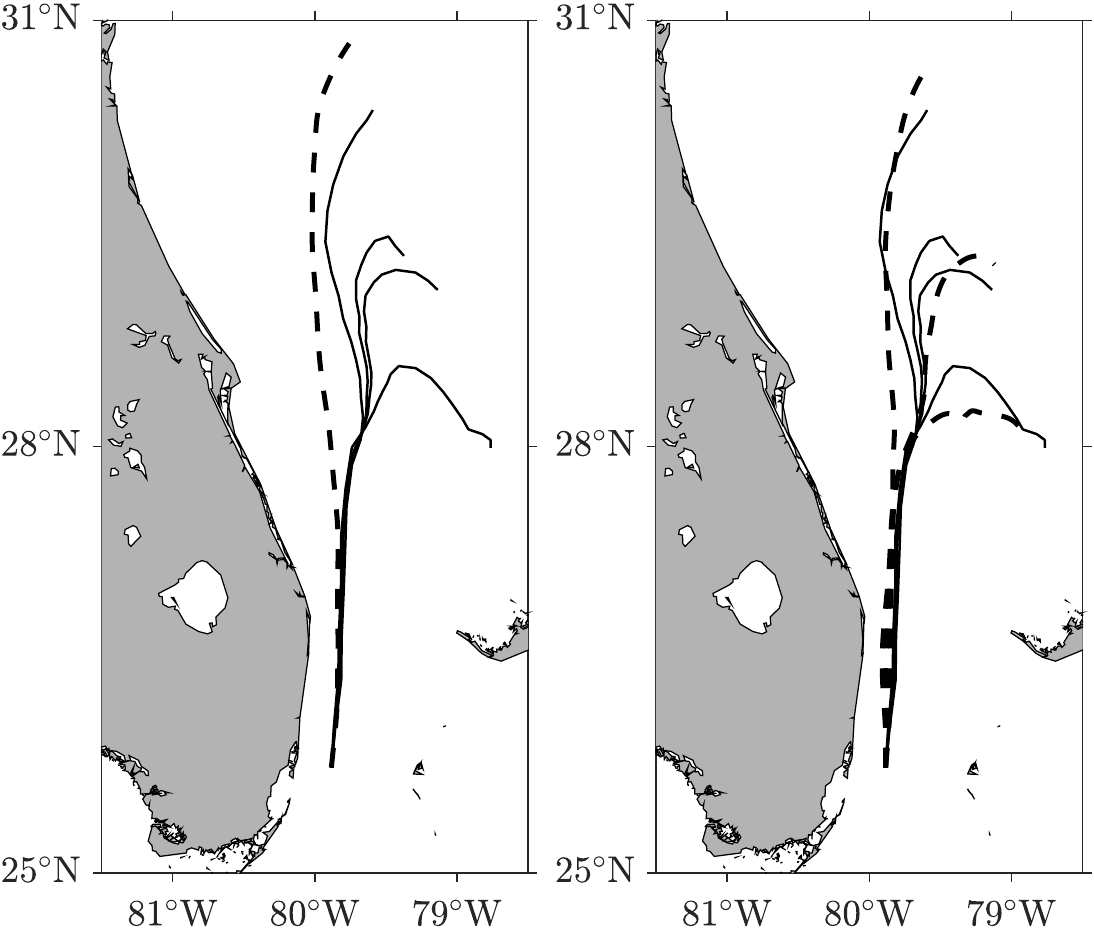}%
  \caption{(left) Trajectories of the special drifters (thin) and
  trajectories resulting from integrating a surface ocean current
  synthesis of altimetry-derived geostrophic flow, wind-induced
  Ekman drift, and drogued drifter velocities (dashed). (right) As
  in the left panel, but with dashed curves resulting from integrating
  leeway velocities constructed by adding to the altimetry/wind/drifter
  velocity synthesis small fractions (from top to bottom 1, 3, and
  5\pct) of wind velocity.}
  \label{fig:aviso-leeway}%
\end{figure}

Four types of special drifters were involved in the experiment.
Three of them were comprised of a main body, made of Styrofoam, and
a small, few-cm-long weighted drogue at the bottom to ensure that
a SPOT\textsuperscript{\textregistered}  trace Global Positioning
System (GPS) tracker was maintained above the sea level.  This
tracker transmitted positions every 6 h.  The main bodies of these
special drifters represented a sphere of radius 12 cm, approximately,
a cube of about 25 cm side, and a cuboid of approximate dimensions
30 cm $\times$ 30 cm $\times$ 10 cm.  These special drifters were
submerged below the sea level by roughly 10, 6.5, and 5 cm,
respectively.  The fourth special drifter, made of plastic, was
designed to mimic a macroalgal mat, such as a \emph{Sargassum} mat.
The GPS tracker was collocated inside a small Styrofoam cone embedded
in the mat.  The maximal area spanned by the plastic mat was of
about 250 cm $\times$ 50 cm and had a thickness of nearly 2 cm.  It
floated on the surface with the majority of its body slightly above
the surface.

In this paper we focus on the analysis of the first week of trajectory
records.   There are two reasons for restricting to this period of
time.  First, the cube stopped transmitting position after one week.
Thus extending the period of analysis beyond one week will shrink
the space of parameters for exploration.  Second, the special
drifters tend to absorb water. This results in a change in their
initial buoyancy over time and thus in their response to ocean
current and wind drag.  In the absence of empirical evidence,
simulating this response will require one to propose some model for
the time variation of the buoyancy, which we avoid to reduce
uncertainties.  With this in mind, we note that the special drifters
were affected by a strong wind event that took place between 2 and
3 days after deployment (Fig.\ \ref{fig:time-wind}, top).  This
wind event unevenly impacted the trajectories, suggesting dominance
of inertial effects. Furthermore, even prior to the anomalous wind
event, the velocity of the special drifters was not uniform across
them (Fig.\ \ref{fig:time-wind}, bottom), suggesting an uneven
response of their motion to the ocean currents as well.  This
reinforces the idea that inertial effects dominated the motion of
the special drifters.

Indeed, surface velocities alone cannot explain the different
trajectories described by the special drifters, as is shown in the
left panel of Fig.\ \ref{fig:aviso-leeway}.  The dashed curve in this
figure is the trajectory that results from integrating a surface
velocity representation starting from the deployment site and time.
The thin curves are the various special drifter trajectories.  The
surface velocity corresponds to a synthesis of geostrophic flow
derived from multisatellite altimetry measurements \citep{LeTraon-etal-98}
and Ekman drift induced by wind from reanalysis \citep{Dee-etal-11},
combined to minimize differences with velocities of GDP drifters
drogued at 15 m \citep{Lumpkin-Pazos-07}.

Moreover, a leeway velocity model is not capable of representing
the variety of trajectories produced by the special drifters with
a single windage strength choice.  Several windage levels must be
considered depending on the special drifter.  This is insinuated
in the right panel of Fig.\ \ref{fig:aviso-leeway}, which shows (in
dashed) trajectories resulting from integrating leeway velocities
constructed by adding to the above surface velocity synthesis small
fractions of the reanalyzed wind field involved in the synthesis.
The windage levels are in the widely used ad-hoc range 1--5\pct.
\cite{Duhec-etal-15, Trinanes-etal-16, Allshouse-etal-17} Which
level best suits a given special drifter cannot be assessed a priori.
The Maxey--Riley theory of \citet{Beron-etal-19b} provides means
for resolving this uncertainty by explicitly accounting for the
effects of the inertia of the drifters on their motion.

\section{The Maxey--Riley framework}\label{sec:mrf}

Consider a stack of two homogeneous fluid layers.  The fluid in the
bottom layer represents the ocean water and has density $\rho$.
The top-layer fluid is much lighter, representing the air; its
density is $\rho_\mathrm{a} \ll \rho$.  Let $\mu$ and $\mu_\mathrm{a}$
stand for dynamic viscosities of water and air, respectively.  The
water and air velocities vary in horizontal position and time, and
are denoted $v(x,t)$ and $v_\mathrm{a}(x,t)$, respectively, where
$x = (x^1,x^2)$ denotes Cartesian \footnote{On the sphere, with
local coordinates $\smash{x^1} = (\lambda-\lambda_0)\cdot
a_\odot\cos\vartheta_0$ and $\smash{x^2} = (\vartheta-\vartheta_0)\cdot
a_\odot$, where $a_\odot$ is the planet's mean radius and $\lambda$
(resp., $\vartheta$) is longitude (resp., latitude), the Maxey--Riley
set takes the form \eqref{eq:MR} with $f = 2\Omega\sin\vartheta +
\tau_\odot v^1$, where $\Omega$ is the planet's rotation rate and
$\tau_\odot := \smash{a_\odot^{-1}}\tan\vartheta$,
$\smash{\frac{\D{}}{\D{t}}}v = \partial_t v + (\nabla v)
\smash{(\gamma_\odot^{-1}v^1,v^2)} = \partial_t v +
\smash{\gamma_\odot^{-1}(\partial_1 v)}v^1 + (\partial_2 v)v^2 $,
where $\gamma_\odot := \sec\vartheta_\odot\cos\vartheta$, and $\omega
= \smash{\gamma_\odot^{-1}}\partial_1v^2 - \partial_2v^1 + \tau_\odot
v^1$. In turn, the reduced Maxey--Riley set takes the form
\eqref{eq:MRslow} with $\smash{\frac{\D{}}{\D{t}}}u = \partial_t u
+ (\nabla u) \smash{(\gamma_\odot^{-1}u^1,u^2)} = \partial_t u +
\smash{\gamma_\odot^{-1}(\partial_1 u)}u^1 + (\partial_2 u)u^2$
with $\gamma_\odot$, and $f$ and $\omega$ as above.} position with
$x^1$ (resp., $x^2$) pointing eastward (resp., northward) and $t$
is time.  This configuration is susceptible to (Kelvin--Helmholtz)
instability \cite{LeBlond-Mysak-78}, which is ignored assuming that
the air--sea interface remains horizontal at all times.  In other
words, any wave-induced Stokes drift \cite{Phillips-77} is accounted
for implicitly by absorbing its effects in the water velocity $v$.
Consider finally a solid spherical particle, of radius $a$ and
density $\rho_\mathrm{p}$, floating at the air--sea interface.
Define \cite{Beron-etal-19b}
\begin{equation}
  \delta := \frac{\rho}{\rho_\mathrm{p}} \ge 1.
\end{equation}
Under certain conditions, clarified in Section \ref{sec:mrdom}, 
$\delta^{-1}$ approximates well the fraction of particle volume
submerged in the water \cite{Beron-etal-16, Beron-etal-19b}.  For
future reference consider the following parameters depending on the
\emph{inertial particle buoyancy} $\delta$:
\begin{equation}
  \Phi := \frac{\mathrm{i}\sqrt{3}}{2}
  \left(\frac{1}{\varphi}-\varphi\right) - \frac{1}{2\varphi} -
  \frac{\varphi}{2} + 1,
  \label{eq:Phi}
\end{equation}
where
\begin{equation}
  \varphi := \sqrt[3]{\mathrm{i}\sqrt{1-(2\delta^{-1}-1)^2} +
  2\delta^{-1} - 1}.
  \label{eq:phi}
\end{equation}
Nominally ranging in the interval $[0,2)$, $\Phi$ allows one to
evaluate the height (resp., depth) of the emerged (resp., submerged)
spherical cap as $\Phi a$ (resp., $(2-\Phi)a$).\cite{Beron-etal-19b}
Finally,
\begin{equation}
  \Psi := \pi^{-1}\cos^{-1}(1-\Phi)
  - \pi^{-1}(1-\Phi) \sqrt{1-(1-\Phi)^2},
  \label{eq:Psi}
\end{equation}
which nominally ranges in $[0,1)$ and gives the emerged (resp.,
submerged) particle's projected (in the flow direction) area as
$\pi\Psi a^2$ (resp., $\pi(1 - \Psi)a^2$).\cite{Beron-etal-19b}

\subsection{The full set}

The Maxey--Riley set \cite{Maxey-Riley-83, Gatignol-83, Auton-etal-88}
includes several forcing terms that describe the motion of solid
spherical particles immersed in the unsteady nonuniform flow of a
homogeneous viscous fluid.  These terms are the \emph{flow force}
exerted on the particle by the undisturbed fluid; the \emph{added
mass force} resulting from part of the fluid moving with the particle;
and the \emph{drag force} caused by the fluid viscosity.

Vertically integrating across the particle's extent the Maxey--Riley
set, enriched by further including the \emph{lift force}
\cite{Montabone-02}, which arises when the particle rotates as it
moves in a (horizontally) sheared flow \cite{Auton-87}, and the
\emph{Coriolis force} \cite{Beron-etal-15, Beron-etal-16,
Haller-etal-16}, which is the only perceptible effect of the planet's
rotation in the $x$-frame (as it has the local vertical sufficiently
tilted toward the nearest pole to counterbalance the centrifugal
force \cite{Gill-82}), Beron-Vera et al.\ \cite{Beron-etal-19b}
obtained the following \emph{Maxey--Riley set for surface ocean inertial
particle motion}:
\begin{equation}
  \dot v_\mathrm{p} + \Big(f +
  \frac{1}{3}R\omega\Big)v_\mathrm{p}^\perp
  + \tau^{-1} v_\mathrm{p} =
  R\frac{\D{v}}{\D{t}} + 
  R\Big(f + \frac{1}{3}\omega\Big)v^\perp +
  \tau^{-1}u,
  \label{eq:MR}
\end{equation}
where
\begin{equation}
  u := (1-\alpha)v + \alpha v_\mathrm{a}. 
  \label{eq:u}
\end{equation}

In \eqref{eq:MR} $v_\mathrm{p}$ is the velocity of the inertial
particle and $\dot v_\mathrm{p}$ its acceleration; $f = f_0 + \beta
x^2$ is the Coriolis parameter; $\omega = - \nabla\cdot v^\perp =
\partial_1v^2 - \partial_2v^1$ is the (vertical component of the)
water's vorticity; $\smash{\frac{\D{}}{\D{t}}}v = \partial_t
v + (\nabla v)v = \partial_t v + (\partial_1v)v^1 + (\partial_2v)v^2$
is the total derivative of the water velocity along an ocean
water particle trajectory;  and parameters
\begin{equation}
  R : = \frac{1 - \frac{1}{2}\Phi}{1 - \frac{1}{6}\Phi} \in [0,1),
  \label{eq:R}
\end{equation}
and
\begin{equation}
  \tau := K\cdot \frac{1-\frac{1}{6}\Phi}{3\left(k^{-1}(1-\Psi) +
  \gamma k_\mathrm{a}^{-1}\Psi\right)\delta
  }\cdot \frac{a^2}{\mu/\rho} > 0,
  \label{eq:tau}
\end{equation}
which measures the \emph{inertial response time} of the medium to
the particle.  The nominal range of $\tau$ values is clarified in
Section \ref{sec:mrdom}.  In \eqref{eq:tau}
\begin{equation}
  \gamma := \frac{\mu_\mathrm{a}}{\mu} > 0;
  \label{eq:gamma}
\end{equation}
parameter $k > 0$ (resp., $k_\mathrm{a}>0$) determines the projected
length scale of the submerged (resp., emerged) inertial particle
piece upon multiplication by the immersion (resp., emersion) depth
(resp., height); and $0 < K \le 1$ is a correction factor that
accounts for the effects of particle's shape deviating from spherical,
satisfying \cite{Ganser-93}
\begin{equation}
  K^{-1} = \frac{1}{3}\frac{a_\mathrm{n}}{a_\mathrm{v}} +
  \frac{2}{3}\frac{a_\mathrm{s}}{a_\mathrm{v}}.
  \label{eq:K}
\end{equation}
Here $a_\mathrm{n}$, $a_\mathrm{s}$, and $a_\mathrm{v}$ are the
radii of the sphere with equivalent projected area, surface
area, and equivalent volume, respectively, whose average provide
an appropriate choice for $a$.  Finally, in \eqref{eq:u}
\begin{equation}
  \alpha := \frac{\gamma k_\mathrm{a}^{-1}\Psi}{k^{-1}(1-\Psi) +
  \gamma k_\mathrm{a}^{-1}\Psi}.
  \label{eq:alpha}
\end{equation}
Since $0\le \alpha  <1$, nominally, the convex combination \eqref{eq:u}
represents a \emph{weighted average of water and air velocities}.

\subsection{Slow manifold approximation}

Set \eqref{eq:MR} represents a nonautonomous four-dimensional
dynamical system in position ($x$) and velocity ($v_\mathrm{p}$).
A two-dimensional system in $x$, which does not require specification
of initial velocity for resolution, can be derived by noting that
\eqref{eq:MR} is valid for sufficiently small particles or,
equivalently, the inertial response time $\tau$ is short enough.
More specifically, \eqref{eq:MR} involves both slow (position) and
fast (velocity) variables, which makes it a singular perturbation
problem.  This enables one to apply geometric singular perturbation
analysis \cite{Fenichel-79, Jones-95} extended to the nonautonomous
case \cite{Haller-Sapsis-08} to obtain \cite{Beron-etal-19b}:
\begin{equation}
  \dot{x} = v_\mathrm{p} = u +  \tau u_\tau
 \label{eq:MRslow}
\end{equation}
$+\,O(\tau^2)$ as $\tau\to 0$, where 
\begin{equation}
  u_\tau :=
  R\frac{\D{v}}{\D{t}} + R \Big(f +
  \frac{1}{3}\omega\Big) v^\perp - \frac{\D{u}}{\D{t}} -
  \Big(f + \frac{1}{3}R\omega\Big) u^\perp
  \label{eq:utau}
\end{equation}
with $\smash{\frac{\D{}}{\D{t}}}u$ being the total derivative of
$u$, defined in \eqref{eq:u}, along a trajectory of $u$.

\begin{figure}[t]
  \linespread{1}\selectfont{}
  \centering%
  \includegraphics[width=.6\textwidth]{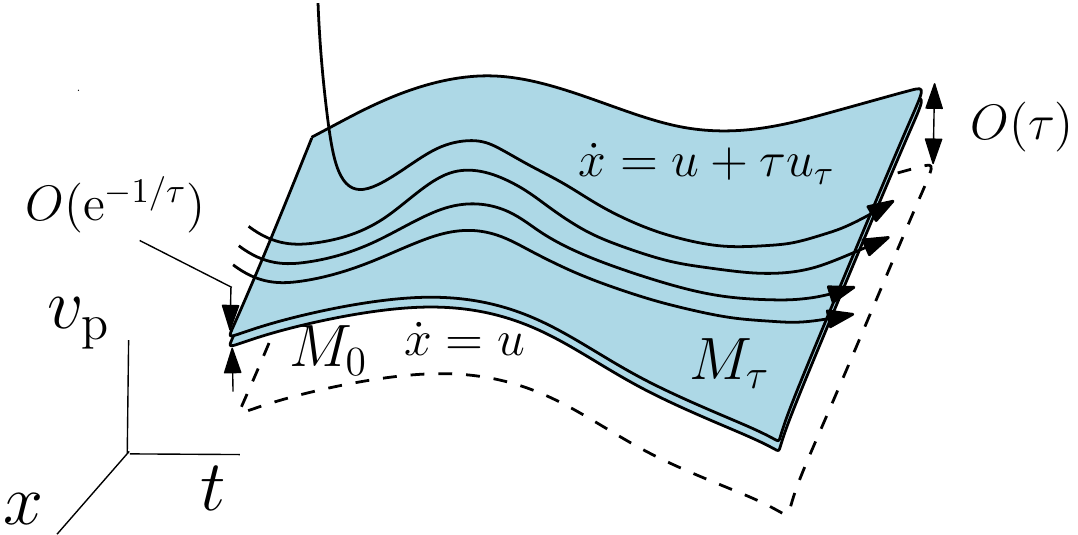}%
  \caption{Geometry of the Maxey--Riley set \eqref{eq:MR} dynamics
  in the extended phase space.  Unique up to an error of
  $O(\mathrm{e}^{-1/\tau}) \ll O(\tau)$, the locally invariant slow
  manifold $M_\tau$ \eqref{eq:Mtau} normally attracts all solutions
  of the Maxey--Riley set when $\tau > 0$ is small exponentially
  fast.  This lies $O(\tau)$-close to the critical manifold $M_0$.
  For the fast dynamics, i.e., with $t$ rescaled by $\tau^{-1}$,
  $M_0$ is filled with fixed points, while for the slow dynamics,
  i.e., with $t$ unscaled, motion on $M_0$ is nontrivial, evolving
  according to the buoyancy-weighted average of water and air
  velocities $u$ \eqref{eq:u}.  Yet motion off $M_0$ is not controlled
  by the dynamics on it.}
  \label{fig:M}%
\end{figure}

The reduced set \eqref{eq:MRslow} controls the evolution of the
full set \eqref{eq:MR} on the manifold 
\begin{equation}
  M_\tau := \{(x,v_\mathrm{p},t) : v_\mathrm{p} = u(x,t) + \tau
  u_\tau(x,t)\},
  \label{eq:Mtau}
\end{equation}
which is referred to as a \emph{slow manifold} because \eqref{eq:MR}
restricted to $M_\tau$, i.e., \eqref{eq:MRslow}, represents a slowly
varying system (Fig.\ \ref{fig:M}).  Invariant up to trajectories
leaving it through its boundary, and unique up to an error of
$O(\mathrm{e}^{-1/\tau}) \ll O(\tau)$,\cite{Fenichel-79} $M_\tau$
normally attracts \emph{all} solutions of the $\tau\to 0$ limit of
\eqref{eq:MR} exponentially fast.  The only caveat \cite{Haller-Sapsis-08}
is that rapid changes in the carrying flow velocity, represented
by $u$, can turn the exponentially dominated convergence of solutions
on $M_\tau$ not necessarily monotonic over finite time.

\section{Clarification of the Maxey--Riley
set}\label{sec:mrclar}

\subsection{Critical manifold}

The $\tau = 0$ limit of \eqref{eq:MR} with $t$ rescaled by $\tau^{-1}$
to form a \emph{fast} timescale has a large set of \emph{fixed
points}, which, given by $v_\mathrm{p} = u$, entirely fill $M_0$,
called the \emph{critical manifold}.  Motion on $M_0$ is thus trivial
for the $\tau = 0$ limit of the fast form of \eqref{eq:MR}.  The
$\tau = 0$ limit of the \emph{slow} form of \eqref{eq:MR}, i.e.,
with $t$ unscaled, blows this motion up to produce nontrivial
behavior on $M_0$, yet leaving the motion undetermined off $M_0$,
which is controlled by $M_\tau$ when $\tau > 0$ small.

The idea that motion on $M_0$ is trivial \cite{Jones-95} must be
understood in the specific dynamical systems sense above and should
not be confused with implying that $\dot x = u$ cannot support rich
dynamics.  Clearly, rich dynamics can even be supported by the
carrying velocity in the original Maxey--Riley model setting with
a single fluid and a finite-size particle either heavier or lighter
than the fluid.  Yet in that case the interest lies in the potentially
much richer dynamics \cite{Cartwright-etal-10} that inertial effects
may produce.  The situation is different in the present case, wherein
the carrying flow ($u$) depends on the buoyancy of the particle,
cf.\ \eqref{eq:u}, and thus has \emph{inertial effects built in}.
Indeed, $u$ is not given a priori as in the standard fluid mechanics
setting \cite{Cartwright-etal-10}. Rather, it follows from vertically
integrating the drag force \citep{Beron-etal-19b}. In other words,
inertial effects are felt by the particle even when $\tau = 0$.  It
turns out, as we will show below, that $\dot x = $u describes the
trajectories of the special drifters over the period analyzed
reasonably well.

It is important to realize that $\dot x = u$ is quite different
than the so-called \emph{leeway model}, i.e., one of the form $\dot
x = v + \varepsilon v_\mathrm{a}$ where $\varepsilon > 0$ is small.
The leeway factor $\varepsilon$ is, as noted above, commonly chosen
in an ad-hoc manner to reduce differences with observations
\cite{Duhec-etal-15, Trinanes-etal-16, Allshouse-etal-17}.  Yet
buoyancy-dependent models for $\varepsilon$ have been proposed in the
literature \cite{Rohrs-etal-12, Nesterov-18}.  But at odds with the
Maxey--Riley approach, these models are obtained by neglecting
inertia and assuming an exact cancellation between water and air
drag forces.

Clearly, one should not expect that the leading-order contribution
to the reduced Maxey--Riley set \eqref{eq:MRslow} be sufficient to
describe all aspects of inertial particle motion in the ocean.
Examples of relevant aspects include clustering at the center of
the subtropical gyres \cite{Beron-etal-16, Beron-etal-19b}, phenomenon
supported on measurements of plastic debris concentration
\cite{Cozar-etal-14} and the analysis of undrogued drifter trajectories
\cite{Beron-etal-16, Beron-etal-19b}, or the role of mesoscale
eddies as attractors or repellers of inertial particles depending
on the polarity of the eddies and the buoyancy of the particles
\cite{Beron-etal-15, Haller-etal-16, Beron-etal-19b} despite the
Lagrangian resilience of their boundaries \cite{Beron-etal-13,
Haller-Beron-13, Beron-etal-18, Beron-etal-19c}, which is also
backed on observations \cite{vanderMheen-etal-19}.  The cited
phenomena, which act on quite different timescales, all require
both $O(1)$ and $O(\tau)$ terms in \eqref{eq:MRslow} for their
description \cite{Beron-etal-15, Haller-etal-16, Beron-etal-16,
Beron-etal-19b} consistent with the slow manifold $M_\tau$ in
\eqref{eq:Mtau}, rather than the critical $M_0$, controlling the
time-asymptotic dynamics of the $\tau\to 0$ limit of the Maxey--Riley
set \eqref{eq:MR}.

\subsection{Domain of validity}\label{sec:mrdom}

Unlike stated in \citet{Beron-etal-19b}, the domain of applicability
of the Maxey--Riley set is \emph{not} extensible to all possible
$\delta$ values, which nominally range in a very large interval
bounded by 1 from below.  Indeed, the fraction of submerged particle
volume\cite{Beron-etal-19b}
\begin{equation}
  \sigma = \frac{1-\delta_\mathrm{a}}{\delta-\delta_\mathrm{a}},
  \label{eq:sga}
\end{equation}
where
\begin{equation}
  1\le \delta \le \frac{\rho}{\rho_\mathrm{a}} \gg 1,\quad
  \frac{\rho_\mathrm{a}}{\rho} 
  \le \delta_\mathrm{a} :=
  \frac{\rho_\mathrm{a}}{\rho_\mathrm{p}} \le 1,
\end{equation}
as static stability (Archimedes' principle) demands, so $0\le \sigma
\le 1$.  Note that $\rho\gg\rho_\mathrm{a}$ implies
$\delta\gg\delta_\mathrm{a}$ and as a result $\sigma \approx
(1-\delta_\mathrm{a})/\delta$, which may be further approximated
by $\delta^{-1}$ if $\delta_\mathrm{a} \ll 1$.  The latter does not
follow from $\rho\ll \rho_\mathrm{a}$ as incorrectly stated in
\citet{Beron-etal-19b}.  It is an assumption which holds \emph{provided
that $\delta$ is not too large}.  This follows from noting that
$\delta_\mathrm{a} \equiv (\rho_\mathrm{a}/\rho)\cdot \delta$.  Thus
inferences made in \citet{Beron-etal-19b} on behavior as $\delta\to
\infty$ are not formally correct and should be ignored.  In particular,
Section IV.B of \citet{Beron-etal-19b} should be omitted, and the
left and middle panels of Fig.\ 2 in that paper, which shows $\alpha$
as a function of $\delta$ over a large range, should be interpreted
with the above clarification in mind.  Also, the formal ranges of
parameters $\Phi$, $\Psi$, and $R$ are smaller than their nominal
ones (stated above).

Currently underway \cite{Beron-etal-19d} is a corrigendum and
addendum to \citet{Beron-etal-19b} where it is shown that the correct
way to formulate the Maxey--Riley set so it is valid for all possible
buoyancy values is by using, instead of $\delta$, the exact fraction
of submerged volume $\sigma$, as given in \eqref{eq:sga}.  In
\citet{Beron-etal-19d} it is shown, for instance, that the $\sigma\to
0$ (equivalently, $\delta\to \infty$) limit is symmetric with respect
to the $\sigma\to 1$ (equivalently, $\delta\to 1$) limit, as it can
be expected.  Also, additional terms, involving air quantities must
be included, both in the full and reduced sets if $\delta$ is allowed
to take values in its full nominal range.  It is important to note,
however, that for the purposes of the present work, which involves
dealing with observed $\delta$ values not exceeding $4$ or so, these
additional terms can be safely neglected and thus is appropriate
to use sets \eqref{eq:MR} or \eqref{eq:MRslow} as presented above.

\subsection{Parameter specification}

In order for the Maxey--Riley parameters to be fully determined by
the carrying fluid system properties and the inertial particle's
characteristics, the projected length factors, $k$ and $k_\mathrm{a}$,
must first be specified. These should depend on how much the sphere
is exposed to the air or immersed in the water to account for the
effect of the air--sea interface (boundary) on the determination
of the drag. With this in mind, we make the following proposition:
\begin{equation}
  k = k_\mathrm{a} = \delta^{-r},\quad r>0.
  \label{eq:k}
\end{equation}
Making $k = k_\mathrm{a}$ guarantees the leeway factor $\alpha$ in
\eqref{eq:alpha} to grow with $\delta$.  This assures the air
component of the carrying flow field to dominate over the water
component as the particle gets exposed to the air.  This is consistent
with making $k = k_\mathrm{a}$ to decay with $\delta$ as this
guarantees the inertial response time $\tau$ in \eqref{eq:tau} to
shorten as the particle gets exposed to the air.  Indeed, ignoring
boundary effects, for a spherical particle that is completely
immersed in the water $\tau = a^2\rho/3\mu$,\cite{Sapsis-Haller-08,
Beron-etal-19b} while $\tau = a^2\rho_\mathrm{a}/3\mu_\mathrm{a}
\equiv (\rho_\mathrm{a}/\rho\gamma)\cdot (a^2\rho/3\mu)$ if the
particle is fully exposed to the air. Using mean density values
$\rho = 1025$ \si{kg.m^{-3}} and $\rho_\mathrm{a} = 1.2$ \si{kg.m^{-3}},
and mean dynamic viscosity values $\mu = 0.001$ \si{kg.m^{-1}s^{-1}}
and $\mu_\mathrm{a} = 1.8 \times 10^{-5}$ \si{kg.m^{-1}s^{-1}}, the
lower bound on $\tau$ is approximately $0.05\cdot (a^2\rho/3\mu)$.
Clearly, with $k = k_\mathrm{a}$ depending on $\delta$ as in
\eqref{eq:k}, $\lim_{\delta\to\infty}\tau = 0$.  But this limit,
as clarified above, is outside the domain of validity of the
Maxey--Riley set \eqref{eq:MR} or its reduced form \eqref{eq:MRslow}.
It turns out that what really matters once the theory is confronted
with observations is that \eqref{eq:k} makes $\tau$ to decay at a
faster rate with increasing $\delta$ than $k = k_\mathrm{a} = 1$,
which corresponds to setting the projected length of the submerged
(resp., emerged) particle piece to be equal to the submerged depth
(resp., emerged height).  In fact, below we show that $r \approx
3$ best fits observations.  In \citet{Beron-etal-19d} we will report
on results aimed at providing a stronger foundation for \eqref{eq:k}
based on direct numerical simulations of low-Reynolds-number flow
around an spherical cap of different heights.  To the best of our
knowledge, a drag coefficient formula for this specific setup is
lacking.  An important aspect that these simulations, in progress
at the time of writing, will account for is the effect of the
boundary on which the spherical cap rests on, which may lead to
changes to the bounds on $\tau$ noted above.

\section{Using the Maxey--Riley framework to explain the
behavior of the special drifters}\label{sec:mrexp}

Table~\ref{tab:parameters} presents our estimates for the parameters
that characterize the special drifters as inertial particles evolving
according to the Maxey--Riley set, in its full \eqref{eq:MR} or
reduced \eqref{eq:MRslow} version.  These are classified into primary
parameters ($a$, $K$, and $\delta$) and secondary parameters
($\alpha$, $R$, and $\tau$), which derive from the primary parameters.

The radius $a$ and shape correction factor $K$ follow from each
special drifter's dimension and shape specification.  In computing
the buoyancy $\delta$ we relied on the estimate of the immersion
depth ($h$) for each special drifter at the Rosenstiel School of
Marine and Atmospheric Science's pier, in Virginia Key (recall
\eqref{eq:Phi} and that $\Phi(\delta) =  2 - h/a$, which specifies
$\delta$).  This estimate does not account for any change in density
from the coast to the deployment site.  Also, the density changes
along the special drifter trajectories, which is ignored in the
analysis.  As one may fairly suspect, our estimates for the mat's
parameters are the most affected by uncertainty due to the configuration
of this special drifter, which is not a solid object as the other
three.

The values of $\alpha$ and $R$ are obtained from \eqref{eq:alpha}
and \eqref{eq:R}, respectively, assuming \eqref{eq:k} and viscosities
set to typical values ($\mu = 0.001$ \si{kg.m^{-1}s^{-1}} and
$\mu_\mathrm{a} = 1.8\times 10^{-5}$ \si{kg.m^{-1}s^{-1}}).
Determining $\tau$ from \eqref{eq:tau} requires one to specify of
the density of the water (for which we used $\rho = 1025$ \si{kg.m^3}),
and the exponent $r$ in \eqref{eq:k}, which is done as follows.

\begin{table}[t!]
  \linespread{1}\selectfont{}
  \centering
  \begin{tabular}{*9c}%
    \hline%
	 special drifter &  \multicolumn{8}{c}{parameter}\\
	 &  \multicolumn{3}{c}{primary} & & &
	 \multicolumn{3}{c}{secondary}\\
	 \cline{2-4}
	 \cline{7-9}
	 \rule{0pt}{3ex}
	  & $a$ [\si{cm}] & $K$ & $\delta$ & & & $\alpha$ & $R$ & $\tau$
	  [\si{d^{-1}}]\\
    sphere &  12 & 1 & 2.7 & & & 0.027 & 0.51 & 0.002\\
    cube  & 16 & 0.96 & 4 & & & 0.042 & 0.42 & 0.001\\
    cuboid & 13 & 0.95 & 2.5 & & & 0.024 & 0.53 & 0.003\\
    mat & 26 & 0.53 & 1.25 & & & 0.005 & 0.79 & 0.031\\
    \hline%
  \end{tabular}%
  \caption{Parameters that characterize the special drifters as
  inertial particles.}%
  \label{tab:parameters}
\end{table}

Let $V$ and $L$ be typical velocity and length scales, respectively.
With these one can form a nondimensional inertial response time
\cite{Beron-etal-19b}
\begin{equation}
  \frac{\tau}{L/V} =
  \frac{K\big(1-\frac{1}{6}\Phi\big)}{3\left(k^{-1}(1-\Psi) + \gamma
  k_\mathrm{a}^{-1}\Psi\right)\delta }\cdot \mathrm{St}, 
  \label{eq:St}
\end{equation}
where 
\begin{equation}
  \mathrm{St} := \left(\frac{a}{L}\right)^2\mathrm{Re},\quad 
  \mathrm{Re} := \frac{VL}{\mu/\rho}
\end{equation}
are Stokes and Reynolds numbers, respectively.  An appropriate
velocity scale is such that $v = O(V)$ while $v_\mathrm{a} =
O(V/\alpha)$.  This makes sense provided that $\alpha$ is small,
which is satisfied for the special drifters.  Taking $V = 1$
\si{m.s^{-1}}, typical at the axis of the Florida Current, and $L
= 50$ \si{km}, a rough measure of the width of the current, one
obtains that $\mathrm{St}$ is order unity at most for the special
drifters.  Assuming that they are spherical so $K = 1$, i.e., $K$
equals upper bound, the nondimensional inertial response time
\eqref{eq:St} is less than unity. This makes using the Maxey--Riley
set to investigate the special drifters' motion defensible, and
further suggests that such motion can be expected to lie close to
its slow manifold if $r > 1$ in \eqref{eq:k}.

\begin{figure}[t]
  \linespread{1}\selectfont{}
  \centering%
  \includegraphics[width=.5\textwidth]{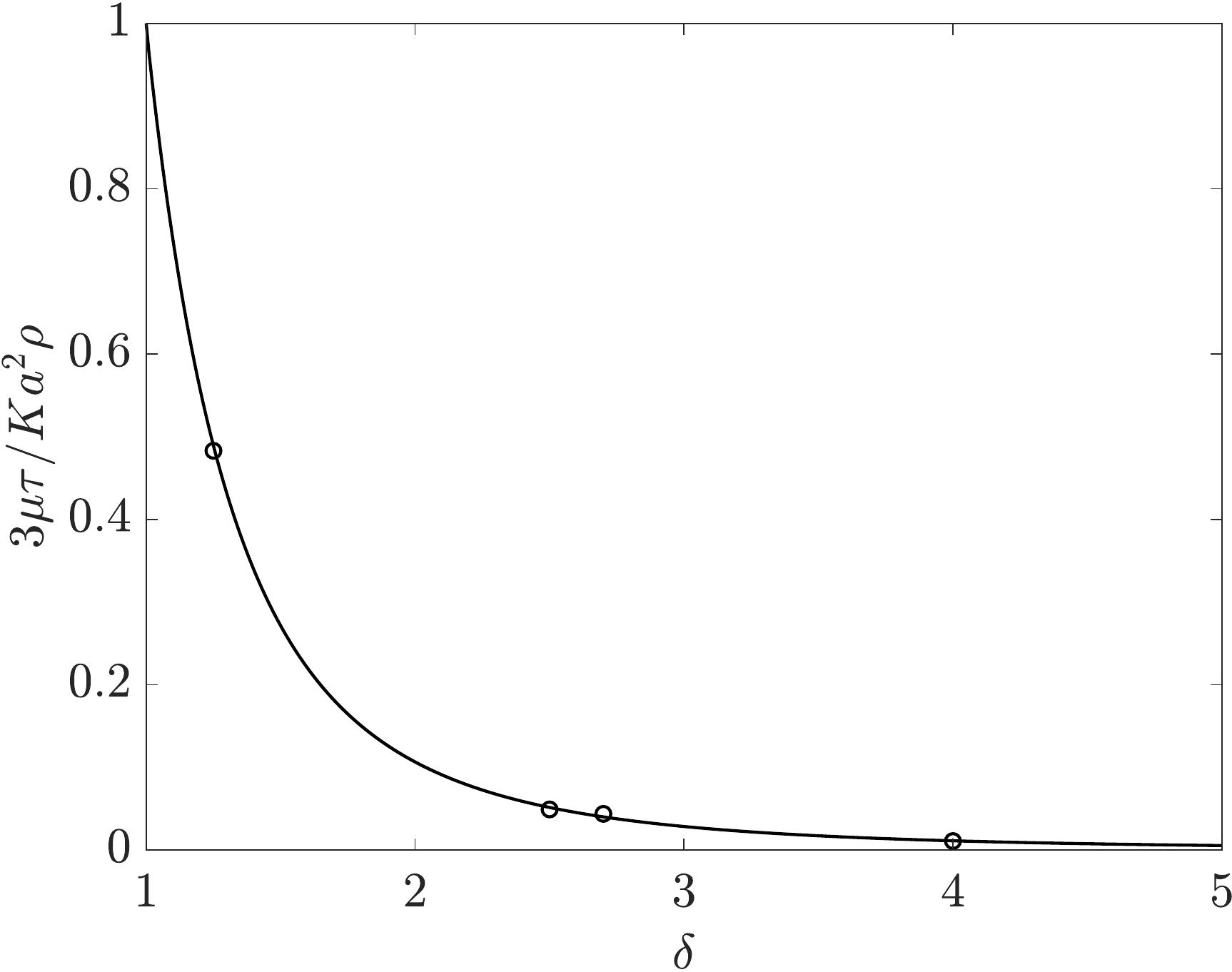}%
  \caption{Optimal inertial response time as a function of special
  drifter buoyancy (circles) and least-squares fit of model
  \eqref{eq:tau-model}.}
  \label{fig:tau}%
\end{figure}

We have estimated the inertial response time $\tau$ that minimizes
the square of the difference between observed special drifter
trajectories and trajectories described by the Maxey--Riley set
\eqref{eq:MR}. The result of this optimization is presented in Fig.\
\ref{fig:tau}, which shows the estimated $\tau$ values (circles)
as a function of special drifter buoyancy ($\delta$).  The curve
is the best fit to a particular $\tau$ model in a least-squares
sense to the optimized $\tau$ values.  The $\tau$ model has one
fitting coefficient given by the exponent ($r$) in the model proposed
for the projected lengths \eqref{eq:k}, namely,
\begin{equation}
  \frac{3\mu}{Ka^2\rho}\cdot \tau(\delta) =
  \frac{1-\frac{1}{6}\Phi(\delta)}{1 + (\gamma-1)\Psi(\delta)}\cdot
  \delta^{-r-1}.
  \label{eq:tau-model}
\end{equation}
Minimization of the square of the residuals gives $r = 2.94$ with
a small one-standard deviation uncertainty ($0.03$) related exclusively
to the goodness of the fit \cite{Ripa-CM-02}.  The optimal values
of $\tau$, which are not different than those resulting using
\eqref{eq:tau-model} with $r = 3$, are listed in
Table~\ref{tab:parameters}.

\begin{figure}[t!]
  \linespread{1}\selectfont{}
  \centering%
  \includegraphics[width=.65\textwidth]{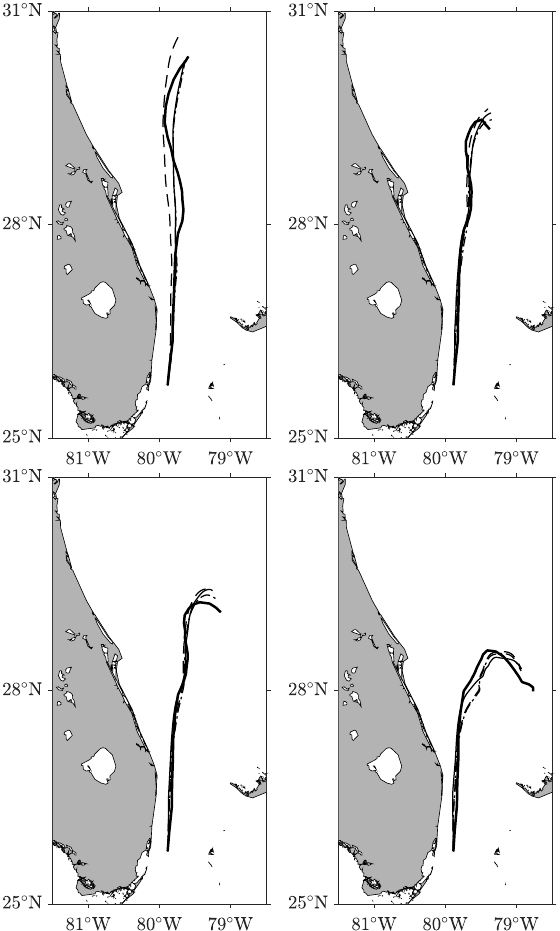}%
  \caption{Special drifter (from left to right, mat, cuboid, sphere,
  and cube) trajectory (solid bold), and trajectories resulting by
  integrating the full Maxey--Riley set (solid thin), the reduced
  Maxey--Riley set (dot-dashed), and the latter with $\tau = 0$
  (dashed).  The water velocity is taken as the surface ocean
  velocity synthesis in Fig.~\ref{fig:aviso-leeway} and the air
  velocity as the reanalyzed wind involved in the synthesis.  Initial
  velocities to integrate the full Maxey--Riley set are taken as
  the velocities of the special drifters. Parameters are given in
  Table \ref{tab:parameters}.}
  \label{fig:mr}%
\end{figure}

With all Maxey--Riley parameters now set, we can proceed to analyze
the trajectories of the special drifters.  In Fig.\ \ref{fig:mr}
we depict special drifter (from left to right, mat, cuboid, sphere,
and cube) trajectories along with trajectories  (solid thin) resulting
by integrating the full Maxey--Riley set \eqref{eq:MR} (solid bold),
trajectories produced by the reduced Maxey--Riley set \eqref{eq:MRslow}
(dot-dashed, nearly indistinguishable from bold solid), and
trajectories resulting by integrating the latter with $\tau = 0$
(dashed).  The surface ocean velocity synthesis discussed in the
preceding section is used to represent the water velocity ($v$)
involved in each of the corresponding dynamical systems, while the
air velocity ($v_\mathrm{a}$) is specified using the reanalyzed
wind data involved in that synthesis.  The initial velocities
required to integrate the full Maxey--Riley set are taken to be
equal to the velocities of the various drifters as obtained from
differentiating their trajectories in time.  Several observations
are in order.

First and foremost is the overall improved agreement between special
drifter trajectories and Maxey--Riley trajectories relative to those
resulting from integrating $v$ and the leeway model $v + \varepsilon
v_\mathrm{a}$ with $\varepsilon = 0.03$ (cf.\ Fig.~\ref{fig:aviso-leeway}).
Indeed, the Maxey--Riley trajectories capture well both the drift
of the mat, predominantly along the Florida Current, and the eastward
turn unevenly experienced by the cuboid, sphere, and cube.  The
leeway model trajectories cannot represent the latter with, as we
note below, a single leeway factor ($\varepsilon$) choice, and the
trajectories of $v$ mainly represent the passive drift of ocean
water along the Florida Current.

A second observation that follows from the inspection of  Fig.\
\ref{fig:mr} is that full Maxey--Riley trajectories coincide,
virtually, with reduced Maxey--Riley trajectories. This indicates
that convergence on the slow manifold is very fast. Consistent with
this is the tendency of the Maxey--Riley trajectories to lie close,
particularly the in the case of the sphere and the cube, to those
produced by the reduced Maxey--Riley set with $\tau = 0$.  This by
no means imply that the special drifters are not affected by inertia.
Quite to the contrary, as we have clarified, $u$ depends on buoyancy
and thus has inertial effects incorporated.  This explains why a
single choice of leeway factor $\varepsilon$ was not sufficient to
explain the uneven effect of the ocean current and wind on the drift
of the special drifters (recall Fig.\ \ref{fig:aviso-leeway}).

An additional observation, which cannot be omitted, is that differences
between observed and Maxey--Riley trajectories, albeit minor compared
with those of the surface velocity synthesis and the leeway model,
are visible in Fig.\ \ref{fig:mr}.  There are several sources of
uncertainty that contribute to produce differences between observed
and Maxey--Riley trajectories.  For instance, there are processes
acting near the surface of the ocean that are not represented by
the surface ocean flow synthesis considered here.  The dominant
component in this synthesis is the altimetry-derived velocity, which
is too coarse to represent submesoscale motions and does not represent
velocity shear between 15-m depth and the ocean surface.  On the
other hand, the Maxey--Riley set, as formulated, can only account
for the potential contribution of wave-induced (Stokes) drift
implicitly, by absorbing the corresponding wave-induced velocity
in the water component of the carrying flow.  The flow synthesis
does not account for wave-induced motions as is constructed in such
a way to minimize differences with velocities of drogued (GDP)
drifters designed to keep wave-induced slip to very low levels
(wind-plus-wave induced slip is less 1 \si{cm.s^{-1}} in 10
\si{m.s^{-1}} wind \cite{Niiler-etal-87}).  In turn, coming from
reanalysis, the near surface wind field cannot be expected to be
fully represented.  There is also uncertainty around the determination
of the buoyancy of the special drifters, which can vary along a
trajectory and this affect its determination even further.

\begin{figure}[t!]
  \linespread{1}\selectfont{}
  \centering%
  \includegraphics[width=.95\textwidth]{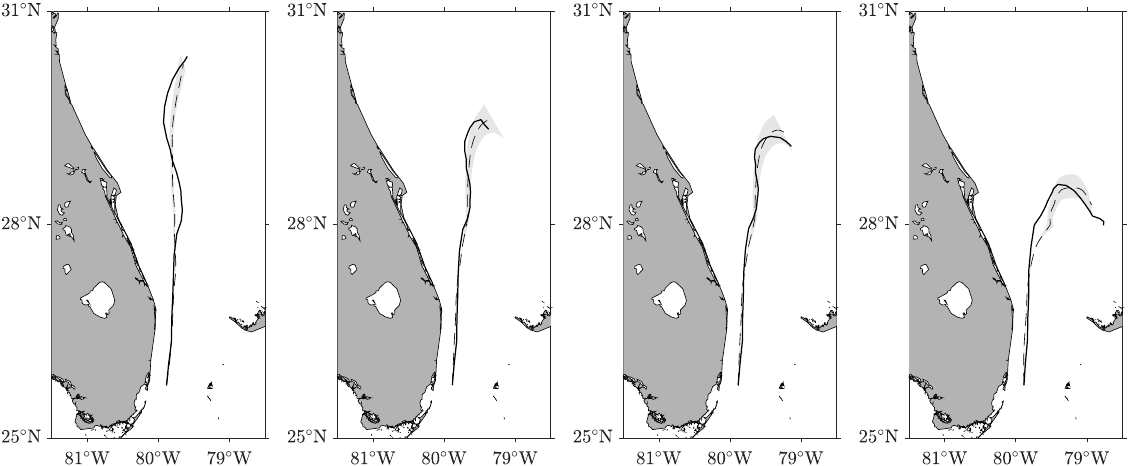}%
  \caption{Special drifter (from left to right, mat, cuboid, sphere,
  and cube) trajectory (solid), Maxey--Riley trajectory with
  parameters as in Table \ref{tab:parameters} (dashed), and area
  spanned by Maxey--Riley trajectories resulting by allowing the
  buoyancy to range in an interval given by the value listed in
  Table \ref{tab:parameters} $\pm10$\pct{} (shaded bands).}
  \label{fig:mr-error}%
\end{figure}

Assessing the effects of the of uncertainty around the determination
of the carrying flow field is not feasible.  Yet we can, at least
roughly, estimate those produced by that around the determination
of the buoyancy of the special drifters.  The result is presented
in Fig.\ \ref{fig:mr-error}, which trajectories (in solid) overlaid
on the area spanned by Maxey--Riley trajectories (shaded bands)
resulting from allowing $\delta$ vary in an interval given by the
value listed in Table \ref{tab:parameters} $\pm10$\pct{} (the dashed
curve, included for reference, has $\delta$ in the center of this
interval).  The width of this $\delta$-interval accounts very roughly
for the error incurred in estimating the submerged depth of the
special drifters in near-coastal water rather than at the deployment
site in the Florida Current, and possibly too any changes in $\delta$
produced by water absorption or ambient water density variations
along trajectories.  Note that the special drifters and corresponding
Maxey--Riley trajectories show consistency among over large portions
to within $\delta$-induced uncertainty.  In particular, most of the
sphere's trajectory falls quite well inside the $\delta$-induced
uncertainty band around the corresponding Maxey--Riley trajectory.
This encourages as to speculate that buoyancy uncertainty dominates
the discrepancies between observed and simulated trajectories.

It is important to realize that differences between observed and
simulated trajectories may never be completely eliminated.  The
fundamental reason for this stands on the unavoidable accumulation
of errors and uncertainties, in addition to sensitive dependence on
initial conditions, in any model, irrespective of how realistic
\cite{Haller-02}.  It is very remarkable then that despite this the
Maxey--Riley set has performed so well when individual trajectories
were compared.

\section{Summary and concluding remarks}\label{sec:con}

In this paper we have presented results of one of a series of
experiments aimed at investigating the mechanism by which objects
floating on the ocean surface are controlled by ocean currents and
winds.  The experiment consisted in deploying simultaneously in the
same location drifting buoys of varied sizes, buoyancies, and shapes
in the Florida Current, off the southeastern Florida Peninsula.
The specially designed drifters described different trajectories,
which were affected by a strong wind event within the first week
of evolution since deployment.  Consistent with the uneven response
to the wind and ocean current action, the differences in the
trajectories were explained as produced by the special drifters'
inertia. This was done by applying a recently proposed Maxey--Riley
theory for inertial (i.e., buoyancy, finite-size) particle motion
in the ocean \cite{Beron-etal-19b}.  Of buoyancy and finite size
effects, the former were found to make the largest contribution to
the inertial effects that controlled the special drifter motion.

The very good agreement between special drifter trajectories and
those produced by the Maxey--Riley may be found surprising given
the uncertainty around the determination of the carrying flow.
Indeed, the ocean component of the flow was provided by a synthesis
dominated by altimetry-derived velocity, while the atmospheric
component was produced by winds from reanalysis.  Both are admittedly
limited.  Furthermore, the Maxey--Riley set does not account for
several potentially important aspects such as space and time
dependence of the particle's buoyancy or wave-induced drift.  

We note that the Maxey--Riley set is found to be similarly successful
in explaining the behavior of special drifters deployed in other
sites of the North Atlantic as part of the experiments that complete
the series.  The drifters have similar characteristics as those
deployed in the Florida Current.  An important difference is that
their trajectories lasted much longer than those discussed here,
resulting in a much more stringent test of the validity of the
Maxey--Riley set.  A detailed analysis is underway and will be
published elsewhere.

Finally, we took the opportunity of this paper to clarify the
Maxey--Riley theory derived in \citet{Beron-etal-19b} with respect
to the nature of the carrying flow and its domain of validity, and
to propose a closure proposal for the determination of the parameters
involved in terms of the carrying fluid system properties and
particle characteristics was proposed.  A corrigendum and addendum
\citet{Beron-etal-19d} to \citet{Beron-etal-19b} is in progress.
This will extend the theory to arbitrary large object's buoyancies
and seek to better justify the closure proposed here by means of
direct numerical simulations.

\begin{acknowledgments}
  The special drifters were constructed by Instrumentation Group's
  personnel Ulises Rivero and Robert Roddy of the National Oceanic
  and Atmospheric Administration's Atlantic Oceanographic and
  Meteorological Laboratory. The altimetry/wind/drifter synthesis
  was produced by RL and can be obtained from ftp:\allowbreak
  //ftp.\allowbreak aoml.\allowbreak noaa.\allowbreak gov/\allowbreak
  phod/\allowbreak pub/\allowbreak lumpkin/\allowbreak decomp.
  Support for this work was provided by the University of Miami's
  Cooperative Institute for Marine \& Atmospheric Studies (MJO,
  FJBV, PM and NFP), National Oceanic and Atmospheric Administration's
  Atlantic Oceanographic and Meteorological Laboratory (JT, RL and
  GJG), and OceanWatch (JT).
\end{acknowledgments}

\bibliography{fot}

\end{document}